\documentclass[12pt]{article}
\usepackage{a4}
\usepackage{epsf}

\newcommand{\beq}{\begin{equation}}
\newcommand{\eeq}{\end{equation}}
\newcommand{\bea}{\begin{eqnarray}}
\newcommand{\eea}{\end{eqnarray}}

\begin{document}

\title{Lattice calculation of $\alpha_s$ in momentum scheme}
\author{ Ph. Boucaud$^a$,  J.P. Leroy$^a$, 
J. Micheli$^a$,   O. P\`ene$^a$,  and C. Roiesnel$^b$ } \par \maketitle
\begin{center}
$^a$Laboratoire de Physique Th\'eorique et Hautes
Energies\footnote{Laboratoire
associ\'e au
Centre National de la Recherche Scientifique - URA D00063}
\\
{Universit\'e de Paris XI, B\^atiment 211, 91405 Orsay Cedex,
France}\\$^b$ Centre de Physique Th\'eorique\footnote{
Unit\'e Mixte de Recherche C7644 du Centre National de 
la Recherche Scientifique\\ 
\\e-mail: Philippe.Boucaud@th.u-psud.fr, roiesnel@cpht.polytechnique.fr
}de l'Ecole Polytechnique\\
F91128 Palaiseau cedex, France 

\end{center}

\begin{abstract}
We compute  the flavorless running coupling constant of QCD from the 
three gluon vertex in the (regularisation
independent) momentum
subtraction renormalisation scheme. This is performed on the 
lattice with high statistics.
 The expected color dependence of the Green functions
 is verified. There are significant $O(a^2\mu^2)$ effects
 which can be consistently controlled. Scaling is demonstrated 
 when the renormalisation scale is varied between 2.1 GeV and 3.85 GeV.
 Scaling when the lattice spacing is varied is very well satisfied. 
The resulting flavorless conventional two loop 
$\Lambda^{(c)}_{\overline {\rm MS}}$ is estimated to be, 
respectively for the MOM and  $\widetilde {\rm MOM}$ scheme,
$361(6)$ MeV and $345(6)$ MeV , while the
three loop results are, depending on $\beta_2$:  
$\Lambda^{(3)}_{\overline {\rm MS}}  = (412 - 59 \frac{\beta_2}
{\beta_{2, \overline{\rm MS}}}
\pm 6\, {\rm MeV}) $ and
$\Lambda^{(3)}_{\overline {\rm MS}}  = (382- 46\frac{\beta_2}
{\beta_{2, \overline{\rm MS}}}\pm
	5\, {\rm MeV}) $.  A preliminary computation of 
	$\beta_2$ in the $\widetilde {\rm MOM}$ scheme leads to
$\Lambda^{(3)}_{\overline {\rm MS}}  = 303(5)\frac {a^{-1}(\beta=6.0)}
{1.97 {\rm GeV}}$ MeV.

\end{abstract}
\begin{flushright} LPTHE Orsay-98/49\\  hep-ph/9810322
\end{flushright}
\newpage

The non-perturbative calculation of the running coupling constant of 
QCD is certainly one very important problem. This program has been performed 
using the Schr\"odinger functional \cite{luscher}, the heavy quark potential
\cite{bali}-\cite{bali2}, the Wilson loop \cite{lepage}, the Polyakov loop
\cite{divitis} and the three gluon coupling \cite{alles}.  The latter method is
the one we will follow in the present letter. 
The principle of the method is quite simple since it consists in following the
steps which are standard in perturbative QCD in the momentum subtraction
scheme. One usually uses as a subtraction point for the three gluon vertex
function an Euclidean point with symmetric momenta:
$p_1^2=p_2^2=p_3^2\equiv\mu^2$.   

In \cite{alles} this non-perturbative minimum subtraction calculation
  has been performed, but using an asymmetric subtraction point 
  $p_1^2=p_3^2\equiv\mu^2,
  p_2=0$. The presence of a vanishing momentum induces some subtleties which
  will be discussed later. The running coupling constant
  computed in \cite{alles} shows a signal of perturbative scaling at large scale. 

In this letter we perform the same program with  symmetric subtraction points,
which is the genuine non-perturbative momentum subtraction  scheme. We also
repeat the work done in \cite{alles}. The whole calculation is achieved
 with a larger statistics, a check of finite volume effects, and a check of 
 scaling when the lattice spacing $a$ is varied. 
 
 In section 1 the general principle of the method is recalled and the systematic construction of the
 symmetric momentum points is summarized. In section 2 the lattice calculation is
 described including the checks of color behaviour, of finite volume effects, 
 and the discussion of 
  $O(a^2\mu^2)$ and three loop effects. Both the scaling in $\mu$ and in $a$
 are demonstrated. 
In section 3 we compare our result for $\Lambda_{\overline{MS}}$ with other
lattice approaches and conclude.

\section{Computing $\alpha_s$ from non-perturbative Green functions}
\label{computing}

In this section we describe the general method used to compute $\alpha_s$ and 
$\Lambda_{\rm
QCD}$ in the continuum, assuming one is able to compute the Euclidean
 Green functions of QCD in the Landau gauge. The lattice aspect will be treated
 in the next section.  
 The principle of the method is exactly the
 standard textbook \cite{politzer} one, generalized to
 non-perturbative QCD \cite{alles}.  

The Euclidean two point Green function in momentum space writes in the
 Landau Gauge: 
\beq
	G_{\mu_1\mu_2}^{(2)\,a_1 a_2}(p,-p)=G^{(2)}(p^2) 
	\delta_{a_1 a_2} \left(\delta_{\mu_1\mu_2}-
	\frac{p_{\mu_1}p_{\mu_2}}{p^2}\right)\label{G2}
\eeq
where $a_1, a_2$ are the color indices ranging from 1 to 8. 

The three-gluon
Green function is equal to the color
tensor\footnote{In general schemes and gauges the $d^{a_1 a_2 a_3}$ tensor should also be
considered, but not in our case as we shall see.} $f^{a_1 a_2 a_3}$ times a momentum dependent function which
 may  be expressed \cite{ball} as
a sum of scalar functions multiplied by tensors: $\sum A_i(p_1^2,p_2^2,p_3^2)
T_{i;\,\mu_1\mu_2\mu_3}$, the $T_{i;\,\mu_1\mu_2\mu_3}$ being built up from
$\delta_{\mu_j\mu_k}$ and momenta. In general there is some arbitrariness in
choosing the tensor basis. One of these may be taken to be the tree level
 three-gluon vertex, projected transversally to the momenta (Landau gauge):
 \beq
 T^{tree}_{\mu_1\mu_2\mu_3}=\big[\delta_{\mu_1'\mu_2'}
 (p_{1}-p_{2})_{\mu_3'}
 + \hbox{cycl. perm.}\big]
 \prod_{i=1,3} \left(\delta_{\mu_i'\mu_i}-\frac {p_{i\,\mu_i'}p_{i\,\mu_i}}
 {p_i^2}\right) \label{tree}
 \eeq 
 while the choice of the other tensors in the tensor basis will be
 explained  below in the particular cases.
 Calling 
 $G^{(3)}(p_1^2,p_2^2,p_3^2)$ the scalar function which multiplies the 
 tensor (\ref{tree}), the renormalised coupling constant in the considered  scheme is 
 given by
 \cite{alles} 
 \beq
 g_R(\mu^2)= \frac{G^{(3)}(p_1^2,p_2^2,p_3^2) Z_3^{3/2}(\mu^2)}
 {G^{(2)}(p_1^2)G^{(2)}(p_2^2)G^{(2)}(p_3^2)}\label{gr}
 \eeq 
where 
\beq
	Z_3(\mu^2)= G^{(2)}(\mu^2) \mu^2\label{Z3}
\eeq
and $\mu^2$ is the renormalisation scale which will be specified in each scheme.

The justification of eq. (\ref{gr}) is standard:
the momentum scheme fixes the renormalisation constants so that the 
two-point and
three-point  renormalised Green functions at the renormalisation point 
take their tree value
with the only substitution of the bare coupling by the renormalised one. 
In particular the renormalised $G^{(2)}_R(p^2)$ takes
its tree value, $1/p^2$, at $p^2=\mu^2$, 
which fixes the
field renormalisation constant (\ref{Z3}). The renormalised coupling
constant is then defined so that the three-point Green function is equal to the bare
tree level one (with the substitution of $g_0$ by $g_R$) at the
symmetric Euclidean point $p_1^2=p_2^2=p_3^2\equiv\mu^2$. 

At the symmetric point only two
independent tensors exist in Landau gauge\footnote{Notice that in the Landau gauge
 the transversality condition with respect to the external momenta,
reduces the number of independant tensors as compared to a general covariant
gauge.}, which we choose to be:
\[
G_{\mu_1\mu_2\mu_3}^{(3)\,a_1 a_2 a_3}(p_1,p_2,p_3)=f^{a_1 a_2 a_3}\Bigg[
G^{(3)}(\mu^2,\mu^2,\mu^2)T^{tree}_{\mu_1\mu_2\mu_3}+\]
\beq
H^{(3)}(\mu^2,\mu^2,\mu^2)\frac{(p_1-p_2)_{\mu_3}
(p_2-p_3)_{\mu_1}(p_3-p_1)_{\mu_2}}{\mu^2}
\Bigg]\label{tensorsym}
\eeq
with $T^{tree}$ defined in (\ref{tree}). To project out
 $G^{(3)}(\mu^2,\mu^2,\mu^2)$ we contract with the appropriate tensor:
 \[
 G^{(3)}(\mu^2,\mu^2,\mu^2)f^{a_1 a_2 a_3}=\frac 1 {18 \mu^2}
  G_{\mu_1\mu_2\mu_3}^{(3)\,a_1 a_2 a_3}(p_1,p_2,p_3)\]
 \beq \left[ T^{tree}_{\mu_1\mu_2\mu_3} + \frac{(p_1-p_2)_{\mu_3}
(p_2-p_3)_{\mu_1}(p_3-p_1)_{\mu_2}}{2 \mu^2}\right]\label{projsym}
 \eeq 
 In the following we will call this momentum configuration the ``symmetric''
 one, and this defines the MOM scheme.
 
We have also considered the $\widetilde{\hbox{MOM}}$ scheme defined by 
subtracting the vertex function  at the asymmetric Euclidean point 
$p_1^2=p_3^2\equiv\mu^2$, $p_2=0$. In Landau gauge there remains only
one tensor, the one in (\ref{tree}), which simplifies:
\beq 
G_{\mu_1\mu_2\mu_3}^{(3)\,a_1 a_2 a_3}(p,0,-p)=2 f^{a_1 a_2 a_3}\,
 p_{\mu_2} 
\left[\delta_{\mu_1\mu_3}-\frac{p_{\mu_1}p_{\mu_3}}{\mu^2}\right]\label{tensorasym}
\, G^{(3)}(\mu^2,0,\mu^2)\eeq
and the scalar factor is extracted via
\beq
G^{(3)}(\mu^2,0,\mu^2)f^{a_1 a_2 a_3}=\frac 1 {6 \mu^2}
  G_{\mu_1 \mu_2 \mu_3}^{(3)\,a_1 a_2 a_3}(p,0,-p)
  \delta_{\mu_1\mu_3}p_{\mu_2}
 \label{projasym}
\eeq
In the following we will call this momentum configuration the ``asymmetric''
 one.
 
Some caution is in order for the latter asymmetric configuration. From (\ref{G2}) it is clear that
\beq
 G^{(2)}(p^2) \delta_{a_1 a_2} = 
 \frac 1 3 \sum_\mu G_{\mu\mu}^{(2)\,a_1 a_2}(p,-p)\label{1/3}
 \eeq
 for any non vanishing value of the momentum. But when the momentum vanishes, 
 the term $p_\mu p_\mu/p^2$ is undetermined. It could seem quite natural to
 follow by continuity formula (\ref{1/3}). On the other hand since only
 the tensor $\delta_{\mu\nu}$ is defined 
 for zero momentum, $G_{\mu\nu}^{(2)}(0,0)=\delta_{\mu\nu} G^{(2)}(0)$
 leads to replacing the 
  factor 1/3 by 1/4. Indeed, the Landau gauge condition does not eliminate
  global gauge transformations, and one additional degree of freedom is left at zero
  momentum.
  This theoretical issue is delicate but
  it is perfectly obvious that the numerical 
 results favor in a dramatic way the factor 1/4. When using the factor 1/3 no 
 sign of perturbative scaling can be seen.
 The factor 1/4  was used in \cite{alles} and we will follow the same recipe.
 
 \subsection{Computing $\Lambda_{\rm QCD}$}
\label{lambdas}

The conventional two-loop $\Lambda^{(c)}$ is obtained in any scheme 
from $\alpha(\mu^2)\equiv  g_R(\mu^2)^2/(4 \pi)$ by   
\beq       
              \Lambda^{(c)} \equiv \mu \exp\left (\frac{-2 \pi}{\beta_0
	      \alpha(\mu^2)}\right)\times
	      \left(\frac{\beta_0  \alpha(\mu^2)}{4 \pi}\right)^{-\frac {\beta_1}
	      { \beta_0^2}}\label{lambda}
\eeq
where
\beq
 \mu\frac{\partial \alpha}{\partial \mu}=-\frac{\beta_0}{2\pi}\alpha^2
 -\frac{\beta_1}{4\pi^2}\alpha^3 -\frac{\beta_2}{64\pi^3}\alpha^4
 -...\label{beta}
\eeq
 $\beta_0=11$, $\beta_1 = 51$ and $\beta_2$ is scheme dependent  ($\beta_{2,
 \overline {\rm MS}}=2857$). Integrating exactly eq. (\ref{beta}) 
 expanded up to order $\alpha^{n+1}$, and imposing the asymptotic limit  
 $\Lambda^{(n)}/\Lambda^{(c)}(\alpha)\to 1$ when $\alpha\to 0$, leads to 
 the definition
 \beq
 \Lambda^{(2)} \equiv \left(1+ \frac{\beta_1 \alpha}{2\pi
 \beta_0}\right)^{\left(\frac{\beta_1}{\beta_0^2}\right)} 
 \Lambda^{(c)}(\alpha)\label{lambda2}
 \eeq
 at two loop and to the three loop $\Lambda^{(3)}$:
 \[
 \Lambda^{(3)}\equiv\Lambda^{(c)}(\alpha)\left(1+\frac {\beta_1\alpha}{2\pi\beta_0}+
 \frac
 {\beta_2\alpha^2}{32\pi^2\beta_0}\right)^{\frac{\beta_1}{2\beta_0^2}}\times\]
 \beq \exp
\left\{\frac{\beta_0\beta_2-4\beta_1^2}{2\beta_0^2\sqrt{\Delta}}\left[
\arctan\left(\frac{\sqrt{\Delta}}{2\beta_1+\beta_2\alpha/4\pi}\right)
-\arctan\left(\frac{\sqrt{\Delta}}{2\beta_1}\right)\right]\right\}
 \label{lambda3}\eeq
 when $\Delta\equiv 2\beta_0\beta_2-4\beta_1^2>0$. 
 
 One simple
 criterium has been proposed 
in \cite{alles} to exhibit perturbative scaling: when plotting (\ref{lambda}) as a
function of $\mu$ perturbative scaling implies that $\Lambda$ should become
constant for large enough $\mu$. We will thus try to fit each of the formulae
(\ref{lambda}), (\ref{lambda2}), (\ref{lambda3}), expressed in terms of 
our measured $\alpha(\mu)$, as a constant. All these formulae converge to the
same $\Lambda$ when $\alpha \to 0$. But since our fits are for $\alpha$ in the
range of 0.3 - 0.5, and since they do not have the same dependence on $\alpha$
they should not all fit our data. 
However, as we shall see, within our errors, acceptable fits
are possible with (\ref{lambda}), and (\ref{lambda3}) varying $\beta_2$ on a
 wide range.  This is due to the fact that, even with our rather large 
 statistics, the three loop effect, 
being only logarithmic, does not modify strongly enough the variation of
$\Lambda(\alpha)$ in our fitting range although the resulting
fitted $\Lambda$ depends significantly on the formula used and on $\beta_2$.
In other words we have different acceptable fits, with slightly different
slopes in $\mu$, which lead asymptotically, when
$\alpha \to 0$, to significantly different $\Lambda$'s. 
There results a systematic error which cannot be fully eliminated until $\beta_2$ is
really computed. 

Notwithstanding this problem, we believe that the possibility to fit several 
of these formulae by a constant on a large range of $\mu$ and with small
statistical errors, 
is an indication that perturbative scaling has been reached. In other words, 
our data show
that the uncertainty is of logarithmic type (higher loops), but there is no room
for significant power corrections. To study power corrections, one has to
consider with care lower scales \cite{burgio}, and we plan to do that in a
forthcoming publication \cite{papier}.

	In order to compare  different schemes, it is standard to 
	translate the results into the $\overline {\rm MS}$ scheme. Once 
	known in any
	scheme,  a one loop computation is enough to yield
	$\Lambda_{\rm QCD}$  in any other scheme 
	\cite{celmaster}, \cite{billoire}.
	 From \cite{celmaster} and
	\cite{alles} we get for zero flavors
	\beq
	\Lambda_{\rm MOM} = 3.334 \Lambda_{\overline {\rm MS}},\qquad
	\Lambda_{\widetilde{\rm MOM}} = \exp(70/66)\simeq 2.888 \Lambda_{\overline {\rm MS}}.
	\eeq

\subsection{Momenta}
\label{momenta}

In a finite hypercubic volume the momenta are the discrete set of vectors 
\beq
p_\mu=2\pi n_\mu/L \label{momentum}
\eeq
where $n_\mu$ are integer  and $L$ is the lattice size. 
The isometry group for momenta is generated by the four
 reflections $p_\mu \to -p_\mu$ for $\mu=1,\cdots,4$ and the permutations between the
 four directions such as $p_x \leftrightarrow p_y $. Altogether this group has
 $2^4 \,4! =384$ elements. We use fully this symmetry in order to increase the
 statistics. The functions   $G^{(2)}(p^2)$ and $G^{(3)}(p^2,0,p^2)$ in eqs
 (\ref{G2}) and (\ref{tensorasym}) are systematically symmetrized over the
 momenta lying on one given group orbit. The number of distinct momenta in an
 orbit is 384 or a divisor of 384 (when the momentum is invariant by some
 subgroup). 

In the case of $G^{(3)}(p^2,p^2,p^2)$ in eq. (\ref{tensorsym}) we furthermore 
symmetrize over the 6 permutations of external legs (Bose symmetry). The number
of elements in an orbit will be $6\times 384=2304$ or one of its divisors.

\subsection{Triplets of external momenta}
\label{triplets} 

We build all triplets of momenta up to some maximum value of the momentum to be
specified later.

In the asymmetric case, there are as many triplets as momenta. For every integer
number there exists at least one orbit of
the isometry group with
$n_\mu n^\mu$ equal to that integer. The number of elements is often a much 
smaller number than 384: 1, 8, 16, 24, 64, etc. 

In the symmetric case, $n_\mu n^\mu$ has to be an {\it even}
number: 
$n_1^2=n_2^2=(n_1+n_2)^2$, where the subindices label the
external particles, and consequently, $n_{1\mu} n_2^{\mu} = - n_1^2/2$ being 
an integer, $n_1^2$ is even. It happens that
for every even integer, we have found at least 
one orbit. The number of elements in one orbit is often a large number, 2304 
and 1152 are frequent, 576 and 192 are
less. Notwithstanding these larger sets, the statistical noise will turn out to be
larger in the symmetric case than in the asymmetric one.   
Let us give some
examples of symmetric triplets: for $n^2=2$:  $(1  1  0  0)( 0  -1  1  0)
(-1  0  -1  0)$ and its 192 tranformed by the isometry-Bose group; for $n^2=4$:
   $(2  0  0  0)(-1  1  1  1)(-1  -1  -1  -1)$ and its 192 transformed. For
   $n^2=16$:   $(4  0  0  0)(-2  2  2  2)(-2  -2  -2  -2)$ its 192 transformed.
   For $n^2=18$ there are 6 orbits, for example   $(4  1  1  0)(-1  -2  -3  2)
   (-3  1  2  -2)$ and its 2304 transformed,  $(3  3  0  0)(-1  -2  3  2)
   (-2  -1  -3  -2)$ and its 1152 transformed, etc.

\section{Lattice calculation of $\alpha_s$ and $\Lambda_{\rm QCD}$.}
\label{lattice}

\subsection{The calculation.}

The calculation has been performed on a QUADRICS QH1, with hypercubic 
lattices of $16^4$ and $24^4$
sites at $\beta=6.0$, $24^4$ at $\beta=6.2$ ( all three with 1000 configurations) 
and $32^4$ at $\beta=6.4$ (100 configurations), combining the Metropolis and the
overrelaxation algorithms. 

The configurations have been transformed to the Landau gauge by a combination of
overrelaxation algorithm and  Fourier acceleration. We end when 
$|\partial_\mu A_\mu|^2 <
10^{-12}$ and when the spatial integral of $A_0$ is constant in time to better
than $10^{-5}$.

We define 
\beq A_\mu(x+ \hat \mu/2) = \frac {U_\mu(x) - U_\mu^\dagger(x)}{2 i a g_0}
- \frac 1 3 \hbox{Tr}\left (\frac{U_\mu(x) - U_\mu^\dagger(x)}
{2 i a g_0}\right ) \label{amu}
\eeq
where $\hat \mu$ indicates the unit lattice vector in the direction $\mu$ and $g_0$ is
the bare coupling constant, and compute the n-point Green functions in momentum space from
\beq
 G^{(n)\,a_1 a_2 \cdots a_n}_{\mu_1\mu_2\cdots\mu_n}(p_1,p_2,\cdots p_n)=
 <A_{\mu_1}^{a_1}(p_1)A_{\mu_2}^{a_2}(p_2)\cdots A_{\mu_n}^{a_n}(p_n)>\label{green}
\eeq
where $p_1+p_2+.. +p_n=0$, $<>$ indicates the Monte-Carlo average
and where
\beq
	A_\mu^a(p)=\frac 1 2 \hbox{Tr}\left [\sum_x A_\mu(x+ \hat \mu/2)
	 \exp(i p (x+ \hat \mu/2))\lambda^a\right]\label{amufour}
\eeq
 $\lambda^a$ being the Gell-Mann matrices and the trace being taken in the
 $3\times 3$ color space.
 
We have computed the Fourier transforms up to a maximum momentum of 
$\simeq 3$ GeV at $\beta=6$, and $\simeq 4.2$ GeV at $\beta=6.2$ 
and $\beta=6.4$. These maxima correspond to $n^2\equiv n_{\mu} n^\mu\le 16$ 
 for $(\beta,V)=(6.0, 16^4)$ and $\le 36$ for all the other
cases.
 
\subsection{Check of the color dependence}
\label{checks}

From the color structure of QCD we expect the two point Green functions to be
proportional to the color tensor $\delta_{a_1 a_2}$. This is indeed the case 
to an accuracy of the order of 1 \%. Furthermore one can prove from gauge 
symmetry (global and local)
and Bose symmetry that the three point Green functions have to be
proportional to $f^{a_1 a_2 a_3}$ in the MOM and $\widetilde
{\rm MOM}$ schemes. This is indeed the case, but the errors now
depend on the momentum. For small values of $n^2$ the
agreement is of a few percent. The errors increase with $n^2$ and when  $n^2$
reaches the $30$'s the error reaches 100 \%.  This is an indication that the
large momenta are grieved by noise. Luckily this caution is necessary only for
the very few largest values of $n^2$ that we have considered. 
Indeed we will exclude the points $n^2>30$ from our fits. In order to reduce the
noise we work 
from now on with color averaged Green functions: $\frac 1 {24} 
G^{(3)\,a_1 a_2 a_3} f^{a_1 a_2 a_3}$ and $\frac 1 {8} 
G^{(2)\,a_1 a_2} \delta_{a_1 a_2 }$.

\begin{figure}
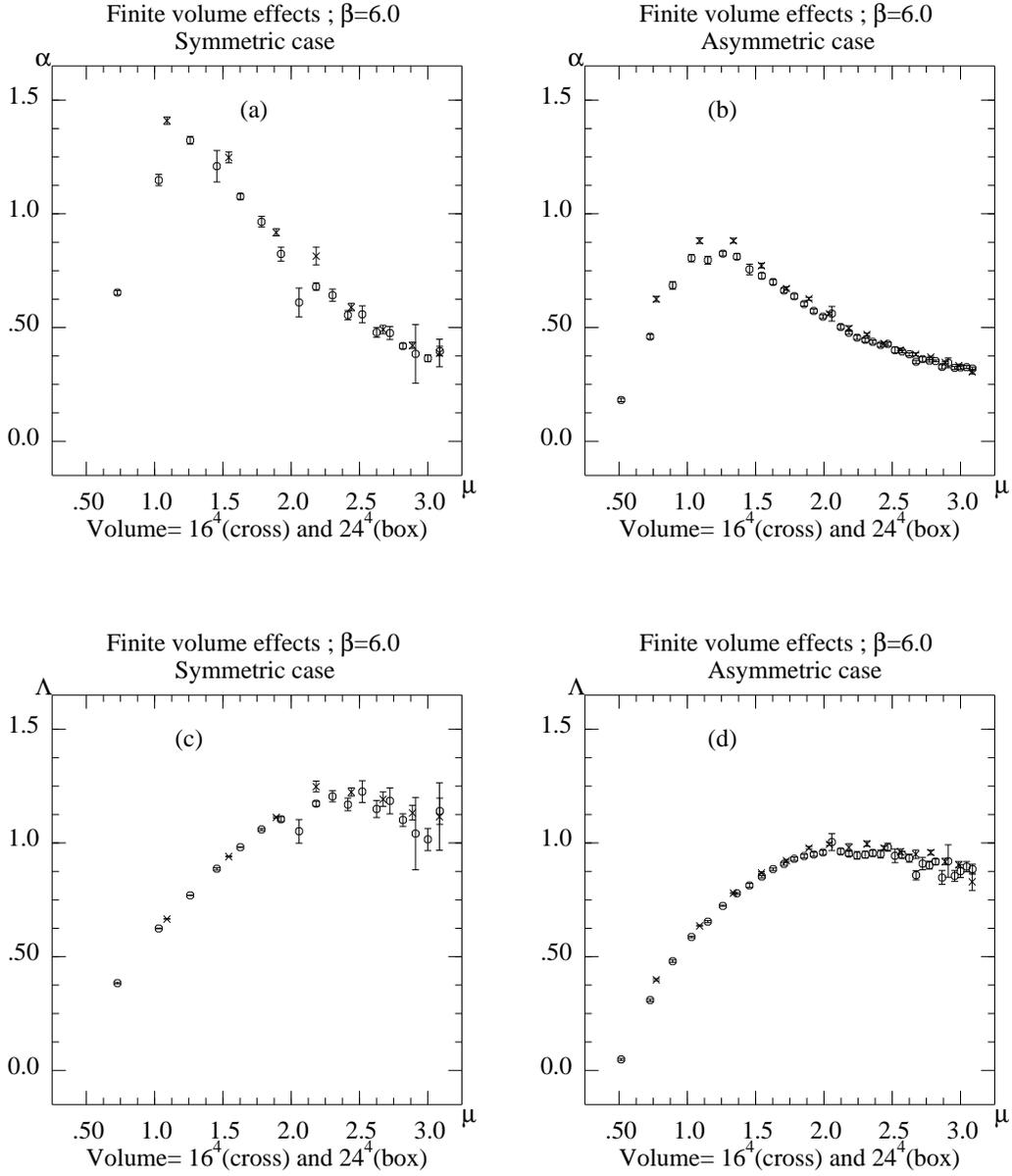

\begin{center}
\leavevmode
\epsfysize=10.0truecm
\epsffile{alpha_sym_vol.eps}
\epsfysize=10.0truecm
\epsffile{alpha_asym_vol.eps}
\\
\vspace*{-1.5cm}
\leavevmode
\epsfysize=10.0truecm
\epsffile{lambda_sym_vol.eps}
\epsfysize=10.0truecm
\epsffile{lambda_asym_vol.eps}
\caption{Comparison  between the  volumes of $16^4$ and $24^4$ at $\beta=6$
for the coupling $\alpha(\mu)$ (figs. a and b), $\Lambda^{(c)}_{\rm MOM}$
(fig. c) and
$\Lambda^{(c)}_{\widetilde {\rm MOM}}$ (fig. d). No ``sinus improvement'' has
 been applied here.}

\label{volumefig}
\end{center}
\end{figure}

\subsection{Finite volume effects}
\label{volume}

The finite volume effects can be checked by a comparison of the two calculations
at $\beta=6.0$, Fig \ref{volumefig}. For relatively small $\mu$, close to the maximum of
$\alpha$, there is a visible decrease 
 of $\alpha$ when the volume is increased. 
 
 For larger $\mu$, the volume dependence
 is still visible, but reduced to a few percent. Comparing the values of
$\Lambda$ fitted in the asymptotic region given in table 1, one finds 
\beq
\Lambda_{sym}(24)/\Lambda_{sym}(16) = 0.96 \pm 0.02, \qquad
\Lambda_{asym}(24)/\Lambda_{asym}(16) = 0.97 \pm 0.02 \label{fini}
\eeq
which indicates that the finite volume effect affects moderately
 the asymptotic estimate of $\Lambda$. 
 More study is needed to quantify precisely this effect. Still,
 from gross estimates, 
    our largest physical volume,  $\beta=6.0,\, 24^4$, lies presumably
  within 5\% above the infinite volume limit.

\subsection{Scaling in $\mu$ and $O(a^2p^2)$ effects}
\label{adeuxpdeux}

\vskip 0.5 cm
\begin{table}
\begin{center}
\begin{tabular}{|c|c|c|c|c|}\hline
$\beta$ & Volume & range (GeV) & $a\Lambda^{(c)}_{\overline {\rm MS}}/(a\sqrt{\sigma})$
sym & $a\Lambda^{(c)}_{\overline {\rm MS}}/(a\sqrt{\sigma})$
asym   \\ \hline
6.0 & $16^4$ & 2.1 - 3.0 & 0.845(25) & 0.809(22) \\
6.0 &  $24^4$ & 2.1 - 3.0 & 0.811(24) & 0.784(23)\\
6.2 & $24^4$ & 2.1 - 3.85 & 0.861(27) & 0.805(21)\\
6.4 & $32^4$ & 2.1 - 3.85 & 0.861(45) & 0.793(27)\\ \hline
\end{tabular}
\vskip 1.5 cm
\caption{Ratios $a\Lambda^{(c)}_{\overline {\rm MS}}/(a\sqrt{\sigma})$ for different
values of $\beta$ after ``sinus improvement'' as explained in section
\ref{adeuxpdeux}. The numbers for $a\sqrt{\sigma}$ are taken from \cite{bali}.
The errors on $a\Lambda$ and $a\sqrt{\sigma}$ have been combined in quadrature.
The scaling invariance is particularly striking especially at constant physical
volume, i.e. comparing lines 1,3,4.  }
\label{tablerat}
\end{center}
\end{table}

\begin{figure}
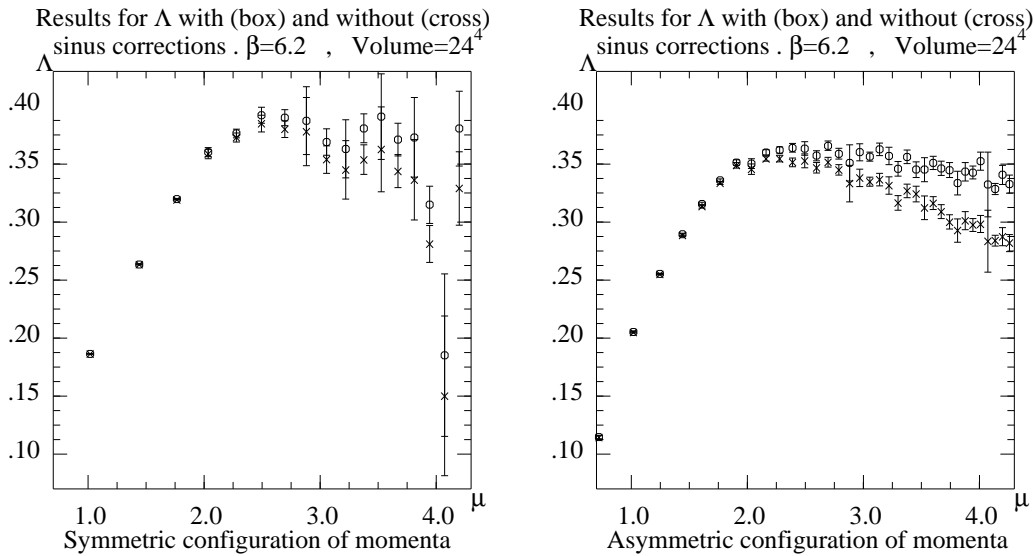

\begin{center}
\leavevmode
\epsfysize=10.0truecm
\epsffile{compar_lambda_MS_sym_6.2_24.eps}
\epsfysize=10.0truecm
\epsffile{compar_lambda_MS_asym_6.2_24.eps}
\vspace*{-3.cm}
\caption{The effect of the ``sinus improvement'' on  
$\Lambda^{(c)}_{\overline {\rm MS}}$  is illustrated in the case of the 1000  configurations at
$(\beta,V)=(6.2,24^4)$. 
A similar improvement can be seen in all cases.}
\label{sinusfig}
\end{center} 
\end{figure}

Figs. \ref{volumefig}(a,b) show the shape of $\alpha(\mu)$. The same shape
is seen for the other $\beta$'s. In fact $\alpha(\mu)$ scales in $a$ to a very good
accuracy. We keep this study for another publication \cite{papier}. 

Turning to the scaling in $\mu$, we see from figs \ref{volumefig}(c,d) 
that both $\Lambda$'s do
 not really show  plateaus at large momentum: they go through a maximum
 around 2 GeV and fall down later on.
   Our study shows that
 this feature cannot be cured simply by a three loop effect. Using eq.
 (\ref{lambda3}) with different values for $\beta_2$ cannot lead to 
  acceptable plateaus for all lattice spacings. Since the fall at large
   $\mu$ is observed systematically, beyond statistical errors,
 but decreases when $\beta$ increases, we conjecture that 
  we deal with an $O(a^2p^2)$ effect. 

We have successfully tried a correction which will be described now. 
 It starts from the remark that in the lattice
Landau gauge, obtained by minimizing $ \sum_{\mu, x} {\rm Tr}[1-U_\mu(x)]$, $p_\mu
A^a_\mu(p)$ does not vanish while $\tilde p_\mu
A^a_\mu(p)$ does, when
$A_\mu^a(p)$ is defined from eq (\ref{amufour}) and where
\beq \tilde p_\mu = \frac 2 a \sin\left(\frac{a p_\mu}{2}\right).\label{ptilde}
\eeq 
The latter momentum differs from the one in (\ref{momentum}) by $O(a^2p^2)$:
$ \tilde p_\mu \simeq p_\mu (1 - \frac 1 {24}a^2 p_\mu^2)$.  It results
 that the  lattice two point Green function is not really  proportional to
the tensor in (\ref{G2}) but to the tensor deduced from (\ref{G2})
with $p_\mu$ substituted by $\tilde p_\mu$, \cite{zwanziger}. 

We perform a similar change in the tensors used to extract $\alpha$. 
The projectors in  (\ref{projsym}) and (\ref{projasym})
 have been normalized to 
give 1 when contracted to the tensors which multiply $G^{(3)}$ in  
(\ref{tensorsym}) and (\ref{tensorasym}) respectively.
 Assuming that the lattice
calculations is such as to produce the tensors in (\ref{tensorsym}) and (\ref{tensorasym})
with $p_\mu$ substituted by $\tilde p_\mu$, there would be a bias in our formulae
(\ref{projsym}) and (\ref{projasym}).
 Indeed the contraction of the 
``tilded'' tensors in (\ref{tensorsym}) and (\ref{tensorasym}) with the tensors
in (\ref{projsym}) and (\ref{projasym}) is smaller than one and decreases with 
increasing $p$. 
We tentatively correct the bias by dividing the result in
(\ref{gr}) by this factor smaller than one. We shall refer to this as the
``sinus improvement''.

For brevity we only show $(\beta,V)=(6.2,\, 24^4)$ in fig. \ref{sinusfig}. The 
improvement of the plateaus is dramatic.
 The large $\mu$ fall has been considerably reduced.
  The improvement is confirmed by a reduction  of
 the $\chi^2$ per degree of freedom 
from exceedingly
large values to acceptable ones, see tables \ref{tableasym} and \ref{tablesym}. 

 Of course, this is only an ad hoc $O(a^2p^2)$
 improvement, by no way  rigorous and systematic. 
Fitting directly a corrective term of the form $1-c a^2 p^2$
leads also to drastically improved $\chi^2$ with best values
of $c$ in  reasonable agreement with the ``sinus improvement'' ($c\sim 1/24$).
  It should be
stressed that the sinus improvement, and the $1-ca^2 p^2$ fits
yield very similar values of $\Lambda$.
We may thus conclude that
 the $O(a^2p^2)$ systematic error on $\Lambda$ is moderate after ``sinus
improvement''.

\vskip 0.5 cm
\begin{table}
\begin{center}
\begin{tabular}{|c|c|c|c|c|}\hline
$\beta$& 6.0&6.0&6.2&6.4\\ \hline
Volume (lat) & $16^4$ & $24^4$ & $24^4$ & $32^4$ \\
Volume (phys) & ($1.63$ fm)$^4$  & ($2.44$ fm)$^4$ & ($1.75$ fm)$^4$ & 
($1.75$ fm)$^4$ \\
range (GeV) & 2.1 - 3.0& 2.1 - 3.0 & 2.1 - 3.85 & 2.1 - 3.85 \\
$\Lambda^{(c)}_{\overline {\rm MS}}$ (MeV) & 356(4)  & 345(6) & 354(5) &
353(11)\\
$\chi^2$/dof & 1.44 & 1.12 & 1.87 & 1.13 \\
$\Lambda^{(2)}_{\overline {\rm MS}}$ (MeV) & 401(5)  & 389(7) & 397(6) &
394(14)\\
 $\chi^2$/dof & 3.13 & 1.66 & 5.99 & 1.43 \\
$\Lambda^{(3)}_{\overline {\rm MS}}$ (MeV) & 314(3) & 303(5) & 313(4) &
312(9)\\
 $\chi^2$/dof & 1.30 & 1.29 & 1.02 & 1.18 \\ 
  unimpr. $\Lambda^{(c)}_{\overline {\rm MS}}$ (MeV) & 332(4) & 324(6) & 338(5) & 
 344(11)\\
  $\chi^2$/dof & 6.51 & 3.71 & 12.3 & 1.58 \\
\hline
\end{tabular}
\vskip 0.5 cm
\caption{Fitted $\Lambda_{\overline {\rm MS}}$ for the {\it asymmetric} 
momentum configurations. The ``sinus improved'' $\Lambda$'s is used except for
 the last two lines. To exhibit the $a$-scaling   we use the ratios 
$\sqrt\sigma_0 (a\Lambda_{\overline {\rm MS}}/(a\sqrt\sigma))$ with 
$\sigma_0=445$ MeV as justified in section \ref{scaling}. $\Lambda^{(3)}$ 
has been computed with $\beta_2=1.69 \beta_{2, 
\overline{\rm MS}}\simeq 4824$. The $\chi^2/$dof's correspond to the fit in the
preceding line.}
\label{tableasym}
\end{center}
\end{table}

\vskip 0.5 cm
\begin{table}
\begin{center}
\begin{tabular}{|c|c|c|c|c|}\hline
$\beta$& 6.0&6.0&6.2&6.4\\ \hline
Volume (lat) & $16^4$ & $24^4$ & $24^4$ & $32^4$ \\
Volume (phys) & ($1.63$ fm)$^4$  & ($2.44$ fm)$^4$ & ($1.75$ fm)$^4$ & 
($1.75$ fm)$^4$ \\
range (GeV) & 2.1 - 3.0& 2.1 - 3.0 & 2.1 - 3.85 & 2.1 - 3.85 \\
$\Lambda^{(c)}_{\overline {\rm MS}}$ (MeV) & 376(6)  & 361(6) & 381(8) &
383(20)\\
 $\chi^2$/dof & 0.38 & 1.06 & 0.89 & 1.50 \\
$\Lambda^{(2)}_{\overline {\rm MS}}$ (MeV) & 442(9)  & 425(10) & 444(12) &
455(30)\\
$\chi^2$/dof & 3.22 & 1.58 & 1.89 & 1.06 \\
$\Lambda^{(3)}_{\overline {\rm MS}}$ (MeV) & 326(5)  & 311(5) & 329(7) &
330(16)\\
 $\chi^2$/dof & 0.19 & 1.25 & 1.12 & 1.79 \\
 unimpr. $\Lambda^{(c)}_{\overline {\rm MS}}$ (MeV) & 364(6) & 351(7) & 372(8) & 
 380(21)\\
 $\chi^2$/dof & 2.99 & 1.49 & 1.69 & 1.24 \\
 \hline

\end{tabular}
\vskip 0.5 cm
\caption{ This table is the analog of table \ref{tableasym} but for the
 {\it symmetric} momentum
configurations.}
\label{tablesym}
\end{center}
\end{table}

\subsection{Three loop effect.}
\label{loop}
A final source of systematic uncertainty comes from our ignorance of 
$\beta_2$ in the MOM scheme. From a preliminary perturbative
calculation \cite{beta2} in the $\widetilde {\rm MOM}$ scheme we get
$\beta_2\simeq 4824$. On the other hand we unsuccessfully tried to fix $\beta_2$ 
non-perturbatively from our asymptotic fits. 
The ratio
$\Lambda^{(3)}/\Lambda^{(c)}$, eqs (\ref{lambda})-(\ref{lambda3}), drops from 1
 when $\alpha$ increases, the drop increasing with $\beta_2$.
  As a result, the fitted value for $\Lambda^{(3)}$ will
 decrease as $\beta_2$ increases. Simultaneously the shape of the plateaus are
  modified. In principle, the requirement of an acceptable
 $\chi^2$ might have restricted the admissible domain for $\beta_2$. Unhappily 
 our preliminary analysis did not turn out to be so restrictive. Only for 
 $\beta_2$ below  $0.5\, \beta_{2, \overline{\rm MS}}$ does the $\chi^2$ become
 prohibitive, see for example the $\beta_2=0$ case ($\Lambda^{(2)}$) in 
 tables \ref{tableasym} and \ref{tablesym}. It might look strange that
 $\Lambda^{(c)}$ fits well while $\Lambda^{(2)}$ does not, both being two-loop
 formulae. In fact $\Lambda^{(c)}$ is only an approximate two-loop 
formula  which can be proven to be 
  very close to $\Lambda^{(3)}$ with 
 $\beta_2 \simeq 8\frac{\beta_1^2}{\beta_0}\simeq 0.66 \beta_{2, \overline{\rm MS}}$.
 
 We therefore cannot do better in the MOM scheme, at present, than to provide  fits
 of $\Lambda^{(3)}$ {\it as a function of} $\beta_2$. For comparison we also provide
 the same analysis in the $\widetilde {\rm MOM}$ scheme. 
 The maximum value for $\beta_2$ which we consider is $\beta_2=2 \beta_{2, 
\overline{\rm MS}}= 5714$ since, for such a large value, the term $\propto
\alpha^4$ in the
$\beta$ function (\ref{beta}) is of the same order as the term $\propto
\alpha^3$ for our range of $\alpha$. If $\beta_2$ was larger than
that, the perturbative expansion would be dubious, and the evidence for perturbative
scaling  shown by our data would appear as a miraculous fake.

\subsection{Scaling in $a$}
\label{scaling}

	The ``sinus improved'' $\Lambda$'s  exhibit a very clear scaling when
	$\mu$ varies above 2.1 GeV, as can be seen from the quality of the plateaus in figs
	\ref{sinusfig} and \ref{grandefig} and from the $\chi^2$ per d.o.f.
	In this subsection we want to study further the scaling when $\beta$,
	i.e. $a$, is varied.  

Since $\Lambda$ depends linearly on $a^{-1}$, the consistency of our fits can
only be checked through a spacing independent ratio. We use the ratios
 $a  \Lambda/(a \sqrt\sigma)$, see table \ref{tablerat},
 where
$a\sqrt\sigma$ is the central value of string tension computed in \cite{bali}. 

In order to write
$\Lambda$ in physical units we then multiply all ratios by one global scale
 factor: $\sqrt\sigma_0=445$ MeV
 tuned  to the central value of a very recent fit \cite{damir} from the $K^\ast$ 
 mass: $a^{-1}{(\beta=6.2)}=2.75 (18)$ GeV. We take the central value: 
 $a^{-1}{(\beta=6.2)}=2.75 $ GeV, whence 
 $a^{-1}{(\beta=6.0)}=1.966 $ GeV and $a^{-1}{(\beta=6.4)}=3.664$ GeV.

	This leads to the plots in fig. \ref{grandefig}. 
	The presence of nice plateaus is
	striking. We fit the average $\Lambda$ on these plateaus, for scales
	never smaller than 2.1 GeV, and as high as allowed by lattice effects. 
The results are presented in tables \ref{tablerat}, \ref{tableasym} and
\ref{tablesym}. The fits for $\Lambda^{(c)}$ and for a large range of
$\Lambda^{(3)}$ yield a $\chi^2$ per
	degree of freedom smaller than 1.5.
	 {\it Scaling in the lattice spacing is striking}, especially for those lattice parameters
	which correspond to a similar physical volume of $\simeq (1.7\, {\rm
	fm})^4$,
	i.e. $(\beta,V)=(6.0,\, 16^4),(6.2,\, 24^4)$ and $(6.4,\, 32^4)$. They
	average to:

\[  \Lambda^{(c)}_{\overline {\rm MS}}  = 378(6)\, {\rm MeV}\ ({\rm symmetric})\qquad 
	\Lambda^{(c)}_{\overline {\rm MS}}  = 355(4)\,{\rm MeV}\ ({\rm asymmetric})
\]
\beq \Lambda^{(3)}_{\overline {\rm MS}}\left(\beta_2=1.69\,\beta_{2\,\overline{\rm MS}}
= 4824\right) \, = \, \left\{\matrix {327(5)\,  {\rm MeV}\
({\rm symmetric})\cr 
	  313(3)\,{\rm MeV}\ ({\rm
	asymmetric})\cr}\right.
\label{resfini}\eeq
where the errors are only statistical. $\beta_2= 4824$ results from our
preliminary calculation \cite{beta2} in the asymmetric scheme.
 For comparison we provide the
result with the same $\beta_2$ in the symmetric case.
The result at the larger volume of $\simeq (2.44 \,{\rm fm})^4$, 
$(\beta,V)=(6.0, 24^4)$, presumably close to the infinite volume limit (section
\ref{volume}), is:
 \[\Lambda^{(c)}_{\overline {\rm MS}}  = 361(6)\, {\rm MeV}\ ({\rm symmetric})\qquad 
	\Lambda^{(c)}_{\overline {\rm MS}}  = 345(6)\, {\rm MeV}\ ({\rm asymmetric})
\]
\beq \Lambda^{(3)}_{\overline {\rm MS}}\left(\beta_2=1.69\, \beta_{2\,\overline{\rm MS}}
= 4824\right) \, = \left\{\matrix {\,311(5)\, {\rm MeV}\ ({\rm symmetric})\cr
 303(5)\, {\rm MeV}\
	 ({\rm asymmetric})\cr}\right.
	\eeq 

Varying $\beta_2$ we find acceptable $\chi^2$'s from $\beta_2 \approx 0.5 \beta_{2, 
\overline{\rm MS}}= 1428$ up
  to beyond $2 \beta_{2, 
\overline{\rm MS}}= 5714$ which we take as the maximum perturbatively consistent
value, see section
\ref{loop}. In this range of $\beta_2$ the fitted $\Lambda^{(3)}$ have,
to a surprisingly good approximation, a linear dependence on $\beta_2$. We
provide the result in the next section.

Finally it is worth mentioning that we have also checked scaling of $\alpha$
 in $a$ over the whole range in $\mu$, including the small values.
  We leave this point for a forthcoming publication
\cite{papier}

\begin{figure}
\begin{center}
\vspace*{-1.0cm}
\leavevmode
\epsfysize=9.0truecm
\epsffile{lambda_MSbar_sym_sin_6.0_16.eps}
\epsfysize=9.0truecm
\epsffile{lambda_MSbar_asym_sin_6.0_16.eps}
\\
\vspace*{-3.cm}
\leavevmode
\epsfysize=9.0truecm
\epsffile{lambda_MSbar_sym_sin_6.0_24.eps}
\epsfysize=9.0truecm
\epsffile{lambda_MSbar_asym_sin_6.0_24.eps}
\\
\vspace*{-3.cm}
\leavevmode
\epsfysize=9.0truecm
\epsffile{lambda_MSbar_sym_sin_6.2_24.eps}
\epsfysize=9.0truecm
\epsffile{lambda_MSbar_asym_sin_6.2_24.eps}
\\
\vspace*{-3.cm}
\leavevmode
\epsfysize=9.0truecm
\epsffile{lambda_MSbar_sym_sin_6.4_32.eps}
\epsfysize=9.0truecm
\epsffile{lambda_MSbar_asym_sin_6.4_32.eps}
\vspace*{-3.cm}
\caption{The fits for $a\Lambda^{(c)}_{\overline {\rm MS}}/(a\sqrt\sigma) 
\sqrt\sigma_0$ (with $\sqrt\sigma_0= 445$  
MeV), including the ``sinus
improvement'' are shown for all studied $\beta$'s and volumes.}
\label{grandefig}
\end{center} 
\end{figure}
\section{Discussions and conclusions}

{\it There is scaling}, as can be seen first from the plateaus of $\Lambda$
as a function of the momentum scale $\mu$, and second from the striking
agreement of the runs for different $\beta$'s. We now quote our final results
from our largest physical volume, $(\beta,V)=(6.0,\, 24^4)$, which we estimate to give
values of $\Lambda$ less than 5\% from the infinite volume limit.

The analysis for symmetric momentum configurations is better grounded
theoretically since it avoids the delicate problem of zero momentum. 
On the other hand, this analysis is noisier than the asymmetric one which
exhibits beautiful plateaus. The good agreement of these two analyses
allows a sort of reciprocal support.
 
Several other lattice estimates of $\Lambda$ have been performed. 
The ALPHA collaboration,
\cite{luscher}, quotes  $\Lambda_{\overline{\rm MS}}=251(21)$ MeV. Other results
are 244(8) MeV
\cite{bali}, $293(18)^{25}_{63}$ MeV \cite{bali2}, 340(50) \cite{alles}. 

Our results for $\Lambda$ happen to be very sensitive to the three loop effect
but $\beta_2$
cannot be fitted non perturbatively from our data. A wide range, $0.5 
\beta_{2, \overline{\rm MS}}=1428 \le \beta_2\le 2
 \beta_{2, \overline{\rm MS}}= 5714$ is allowed, in which
 the three loop $\Lambda^{(3)}$
can be approximated by the following formulae: 
\[ \Lambda^{(3)}_{\overline {\rm MS}}  = \left[\left(412 - 59 \frac{\beta_2}
{\beta_{2, \overline{\rm MS}}}
\pm 6\right)\, {\rm MeV}\right]\frac {a^{-1}(\beta=6.0)}{1.97 {\rm GeV}}\ ({\rm symmetric})\]
\beq 
	\Lambda^{(3)}_{\overline {\rm MS}}  = \left[\left(382- 46\frac{\beta_2}
{\beta_{2, \overline{\rm MS}}}\pm
	5\right)\, {\rm MeV}\right] \frac {a^{-1}(\beta=6.0)}{1.97 {\rm GeV}}\ ({\rm asymmetric})
	\label{resbeta2}\eeq

Comparing the results in both schemes  seems to indicate that
the $\beta_2$'s  in MOM and $\widetilde {\rm MOM}$ schemes
are not too different. A calculation of $\beta_2$ in the MOM scheme would
be most welcome. 

 Our preliminary computation of $\beta_2$ in the   $\widetilde {\rm MOM}$ scheme,
\cite{beta2},  uses the results of \cite{davyd} and yields a value of 
$\beta_{2\,\widetilde {\rm MOM}}\simeq 1.69\,\beta_{2, 
\overline{\rm MS}}\simeq 4824$. {\it Our final result is then}
\beq
\Lambda^{(3)}_{\overline {\rm MS}}= \left(303\pm 5\, {\rm MeV}\right)
 \frac {a^{-1}(\beta=6.0)}{1.97 {\rm GeV}}\quad ({\rm asymmetric})
\eeq

\section*{Acknowledgements.}

These calculations were performed on the QUADRICS QH1 located in the Centre de
Ressources Informatiques (Paris-sud, Orsay) and purchased thanks
to a funding from the Minist\`ere de l'Education Nationale and the CNRS.
We are specially indebted to Francesco Di Renzo, Claudio Parrinello and Carlotta
Pittori for thorough discussions which helped initiating this work
We acknowledge Damir Becirevic, Konstantin Chetyrkin,  Yuri Dokshitzer, 
Ulrich Ellwanger, Gregory Korchemsky and  Alfred Mueller 
 for several inspiring comments.

\end{document}